\documentclass[twoside]{article}
\usepackage{amsmath,amssymb,amsfonts}
\usepackage{qic,epsfig}

\begin{document}
\setlength{\textheight}{8.0truein}    


\normalsize\textlineskip
\thispagestyle{empty}
\setcounter{page}{1}



\alphfootnote

\fpage{1}

\centerline{\bf QUANTUM INFORMATION SPLITTING USING A PAIR OF {\it GHZ} STATES}
\vspace*{0.037truein}

\centerline{\footnotesize
KAUSHIK NANDI}
\vspace*{0.015truein}
\centerline{\footnotesize\it 
Department of Physics,
Gurudas College}
\baselineskip=10pt
\centerline{\footnotesize\it 
Kolkata 700 054, India}
\baselineskip=10pt
\centerline{\footnotesize\it 
Email: mail.knandi@gmail.com}
\vspace*{10pt}
\centerline{\footnotesize
GOUTAM PAUL\footnote{{\em Corresponding author}}}
\vspace*{0.015truein}
\centerline{\footnotesize\it 
Cryptology and Security Research Unit}
\baselineskip=10pt
\centerline{\footnotesize\it 
R. C. Bose Centre for Cryptology and Security}
\baselineskip=10pt
\centerline{\footnotesize\it 
Indian Statistical Institute}
\baselineskip=10pt
\centerline{\footnotesize\it 
203 B. T. Road, Kolkata 700 108, India,
}
\baselineskip=10pt
\centerline{\footnotesize\it 
Email: goutam.paul@isical.ac.in}

\vspace*{0.21truein}

\abstracts{
We describe a protocol for quantum information splitting (QIS) of a restricted class of three-qubit states among three parties Alice, Bob and Charlie, using a pair of GHZ states as the quantum channel. There are two different forms of this three-qubit state that is used for QIS depending on the distribution of the particles among the three parties. There is also a special type of four-qubit state that can be used for QIS using the above channel. We explicitly construct the quantum channel, Alice's measurement basis and the analytic form of the unitary operations required by the receiver for such a purpose. 
}{}{}

\vspace*{10pt}


\vspace*{1pt}\textlineskip    

\section{{\label{sec:intro}}Introduction}
The recent advancements in quantum processing tasks viz. quantum teleportation, quantum cryptography and quantum computation are mainly based on quantum entanglement. Quantum teleportation is a quantum process in which Alice sends a qubit to a distant receiver Bob through the shared entangled channel. This process of transmission and recreation of a qubit state is possible not only in the laboratory setup but in principle over an arbitrary distance. Quantum Information Splitting (QIS) is the process for splitting information through quantum means into several parts so that no split part or parts is sufficient to recover the information but the entire set is. In case of QIS, the information can be classical or quantum in nature. The idea of teleportation has been successfully applied in the process of quantum information splitting. Since a quantum state is perfectly reconstructed with the help of other parties, the process of quantum teleportation involving more than two parties often referred in literature as quantum state sharing (QSTS). 

Bennett {\it et al.}~\cite {BB93} first proposed the idea of quantum teleportation of an arbitrary single qubit using an EPR state as quantum channel. Thereafter, quantum teleportation of a single qubit state was proposed using various multipartite quantum systems. Karlsson {\it et al.}~\cite{KB98} used a tripartite GHZ state for this purpose. Pati had demonstrated a protocol for quantum teleportation using a four-partite GHZ state~\cite{P00} and an asymmetric W state~\cite{AP06}. The perfect teleportation of an arbitrary two-qubit state was first proposed by Rigolin~\cite{R05} using two Bell states. Thereafter, Yeo {\it et al.}~\cite{YC06} had shown that tensor product of two orthogonal states can also be used for this purpose. A genuinely entangled five qubit state was employed by Panigrahi {\it et al.}~\cite{P08} to demonstrate the process of perfect teleportation of arbitrary two qubit states. Two qubit teleportation was also shown by Nie {\it et al.} using five qubit cluster state~\cite{NL10} and six qubit genuinely entangled states~\cite{NJ10}.

The idea of QIS of a single qubit was first due to Hilery {\it et al.}~\cite{HBB99} using three-qubit GHZ states. Later this process was investigated by Karlsson {\it et al.}~\cite{KK99} using three particle entanglement. Cleve {\it et al.}~\cite{C99} used a process similar to error correction and Zheng used W state~\cite{Z06} for the QIS of a single qubit. The QIS of an arbitrary two-qubit state was proposed by Deng {\it et al.} using two GHZ states~\cite{D05}. QIS of two qubit states using cluster states was demonstrated by Nie~\cite{NS11}, Panigrahi~\cite{PM11,PS11} and Han~\cite{H12}. Recently, two-qubit QIS was discussed using arbitrary pure or mixed resource states~\cite{ZL11} and asymmetric multi-particle state~\cite{Z12}.

The organization of this paper is as follows. In Sec.~\ref{sec:3QIS}, we discuss the QIS of two non-equivalent forms of special three-qubit states using a pair of GHZ states as quantum channel. We describe the protocol of QIS of special type of four-qubit states in Sec.~\ref{sec:4QIS}.

\section{\label{sec:3QIS}Quantum Information Splitting of a Three-qubit State}
Any arbitrary three-qubit state cannot be used for QIS using a pair of GHZ states. It is only possible to split a special type of three-qubit state through this type of quantum channel. 

Suppose, Alice, Bob and Charlie share a pair of GHZ states, $|\Psi^0_{\text{GHZ}}\rangle=\frac{1}{\sqrt 2}[|000\rangle + |111\rangle]$ among themselves, and they decide that Bob will get the state. Of the first $|\Psi_{\text{GHZ}}^0\rangle$ state, particle 1 belongs to Alice, Bob has particle 2 and Charlie has particle 3. For the the second $|\Psi_{\text{GHZ}}^0\rangle$ state, Alice has particle 1 and rest belongs to Bob. The quantum channel in the $|AABBBC\rangle$ ($A\to$ Alice, $B \to$ Bob and $C\to$ Charlie) format is then given by
\begin{equation*}
|\Psi_G\rangle =\frac{1}{2}[|000000\rangle+|010110\rangle+|101001\rangle+|111111\rangle].
\end{equation*}
In this case, a special three-qubit state $|\xi\rangle=\alpha|000\rangle+\beta|011\rangle+\gamma|100\rangle+\delta|111\rangle$ in Alice's possession, with $|\alpha|^2+|\beta|^2+|\gamma|^2+|\delta|^2=1$, can be used for QIS. The combined state may be written as
{\small
\begin{equation*}
|\xi\rangle\otimes|\Psi_G\rangle = \frac{1}{2}[\alpha|000\rangle+\beta|011\rangle+\gamma|100\rangle+\delta|111\rangle] \otimes 
[|000000\rangle+|010110\rangle+|101001\rangle+|111111\rangle].
\end{equation*}
}
Alice then measures the five qubits (the first five in the above representation) in her possession in the following orthonormal basis.
In order to teleport the above mentioned restricted class of three-qubit state, the following subset of 16 states (out of 32), denoted by
$\eta_i$, where $i = 0, \ldots 15$, is sufficient. 

\begin{equation*}
|\eta_{i}\rangle  = \frac{1}{2}(\sigma_x^{4})^{i_3}(\sigma_x^{5})^{i_2}[|00000\rangle + (-1)^{i_1} |01101\rangle
+(-1)^{i_0} |10010\rangle + (-)^{i_1+i_0}|11111\rangle],
\end{equation*}
where $i_3i_2i_1i_0$ is the 4-bit binary representation of $i$ and
$\sigma_x^b$ denotes $\sigma_x$ operated on the $b$-th bit ($b=1, \ldots, 5$)
from the left end corresponding to each term. Alternatively, starting with
\begin{equation*}
|\eta_{0}\rangle=\frac{1}{2}[|00000\rangle+|01101\rangle+|10010\rangle+|11111\rangle],
\end{equation*}
we can represent the other states in terms of $|\eta_{0}\rangle$ as
\begin{equation*}
|\eta_{i}\rangle = (\sigma_x^{4})^{i_3}(\sigma_x^{5})^{i_2} (\sigma_z^{4})^{i_1}(\sigma_z^{5})^{i_0}) |\eta_{0}\rangle.
\end{equation*}

After the measurement, Alice communicates her measurement results via four classical bits to Bob. Charlie now measures his qubit in the Hadamard basis and communicates his result to Bob through one cbit. Alice and Charlie's measurement results, their communicated results to Bob and Bob's corresponding operations are listed in Table~\ref{tab1}. Note that Alice's information corresponding
to the state $\eta_i$ is given by the 4-bit binary representation of $i$.
Moreover, Charlie's information is 0 or 1 depending on whether his state is
$|+\rangle$ or $|-\rangle$.

\begin{table}[hb]
\tcaption{Strategy for recovering the three-qubit state}
\centerline{\footnotesize\smalllineskip
\begin{tabular}{c c c c}\\
\hline
Alice's   	&Charlie's 	&Bob's	&Bob's\\
state 	 	&state 		&State 	&Operation\\
\hline
$|\eta_0\rangle$ 	 &$|+\rangle$	&$\alpha|000\rangle+\beta|011\rangle+\gamma|100\rangle+\delta|111\rangle$ &$I \otimes I\otimes I$ \\
\hline
$|\eta_0\rangle$ 	 &$|-\rangle$	&$\alpha|000\rangle+\beta|011\rangle-\gamma|100\rangle-\delta|111\rangle$ &$\sigma_z \otimes I\otimes I$ \\
\hline
$|\eta_1\rangle$ 	 &$|+\rangle$	&$\alpha|000\rangle-\beta|011\rangle+\gamma|100\rangle-\delta|111\rangle$ &$I \otimes \sigma_z\otimes I $ \\
\hline
$|\eta_1\rangle$ 	 &$|-\rangle$	&$\alpha|000\rangle-\beta|011\rangle-\gamma|100\rangle+\delta|111\rangle$ &$\sigma_z \otimes \sigma_z\otimes I $ \\
\hline
$|\eta_2\rangle$ 	 &$|+\rangle$	&$\alpha|000\rangle+\beta|011\rangle-\gamma|100\rangle-\delta|111\rangle$ &$\sigma_z \otimes I \otimes I$ \\
\hline
$|\eta_2\rangle$ 	 &$|-\rangle$	&$\alpha|000\rangle+\beta|011\rangle+\gamma|100\rangle+\delta|111\rangle$ &$I \otimes I \otimes I$ \\
\hline
$|\eta_3\rangle$ 	 &$|+\rangle$	&$\alpha|000\rangle-\beta|011\rangle-\gamma|100\rangle+\delta|111\rangle$ &$\sigma_z \otimes \sigma_z \otimes I$ \\
\hline
$|\eta_3\rangle$ 	 &$|-\rangle$	&$\alpha|000\rangle-\beta|011\rangle+\gamma|100\rangle-\delta|111\rangle$ &$I \otimes \sigma_z \otimes I$ \\
\hline
$|\eta_4\rangle$ 	 &$|+\rangle$	&$\alpha|011\rangle+\beta|000\rangle+\gamma|111\rangle+\delta|100\rangle$ &$I \otimes \sigma_x \otimes \sigma_x $ \\
\hline
$|\eta_4\rangle$ 	 &$|-\rangle$	&$\alpha|011\rangle+\beta|000\rangle-\gamma|111\rangle-\delta|100\rangle$ &$\sigma_z \otimes \sigma_x \otimes \sigma_x $ \\
\hline
$|\eta_5\rangle$ 	 &$|+\rangle$	&$\alpha|011\rangle-\beta|000\rangle+\gamma|111\rangle-\delta|100\rangle$ &$I \otimes i\sigma_y \otimes \sigma_x $ \\
\hline
$|\eta_5\rangle$ 	 &$|-\rangle$	&$\alpha|011\rangle-\beta|000\rangle-\gamma|111\rangle+\delta|100\rangle$ &$\sigma_z \otimes i\sigma_y \otimes \sigma_x $ \\
\hline
$|\eta_6\rangle$ 	 &$|+\rangle$	&$\alpha|011\rangle+\beta|000\rangle-\gamma|111\rangle-\delta|100\rangle$ &$\sigma_z \otimes \sigma_x \otimes \sigma_x $ \\
\hline
$|\eta_6\rangle$ 	 &$|-\rangle$	&$\alpha|011\rangle+\beta|000\rangle+\gamma|111\rangle+\delta|100\rangle$ &$I \otimes \sigma_x \otimes \sigma_x $ \\
\hline
$|\eta_7\rangle$ 	 &$|+\rangle$	&$\alpha|011\rangle-\beta|000\rangle-\gamma|111\rangle+\delta|100\rangle$ &$\sigma_z \otimes i\sigma_y \otimes \sigma_x $ \\
\hline
$|\eta_7\rangle$ 	 &$|-\rangle$	&$\alpha|011\rangle-\beta|000\rangle+\gamma|111\rangle-\delta|100\rangle$ &$I \otimes i\sigma_y \otimes \sigma_x $ \\
\hline
$|\eta_8\rangle$ 	 &$|+\rangle$	&$\alpha|100\rangle+\beta|111\rangle+\gamma|000\rangle+\delta|011\rangle$ &$\sigma_x \otimes I \otimes I$ \\
\hline
$|\eta_8\rangle$ 	 &$|-\rangle$	&$\alpha|100\rangle+\beta|111\rangle-\gamma|000\rangle-\delta|011\rangle$ &$i\sigma_y \otimes I \otimes I$ \\
\hline
$|\eta_9\rangle$ 	 &$|+\rangle$	&$\alpha|100\rangle-\beta|111\rangle+\gamma|000\rangle-\delta|011\rangle$ &$\sigma_x \otimes \sigma_z \otimes I$ \\
\hline
$|\eta_9\rangle$ 	 &$|-\rangle$	&$\alpha|100\rangle-\beta|111\rangle-\gamma|000\rangle+\delta|011\rangle$ &$i\sigma_y \otimes \sigma_z \otimes I$ \\
\hline
$|\eta_{10}\rangle$ 	 &$|+\rangle$	&$\alpha|100\rangle+\beta|111\rangle-\gamma|000\rangle-\delta|011\rangle$ &$i\sigma_y \otimes I \otimes I$ \\
\hline
$|\eta_{10}\rangle$ 	 &$|-\rangle$	&$\alpha|100\rangle+\beta|111\rangle+\gamma|000\rangle+\delta|011\rangle$ &$\sigma_x \otimes I \otimes I$ \\
\hline
$|\eta_{11}\rangle$ 	 &$|+\rangle$	&$\alpha|100\rangle-\beta|111\rangle-\gamma|000\rangle+\delta|011\rangle$ &$i\sigma_y \otimes \sigma_z \otimes I$ \\
\hline
$|\eta_{11}\rangle$ 	 &$|-\rangle$	&$\alpha|100\rangle-\beta|111\rangle+\gamma|000\rangle-\delta|011\rangle$ &$\sigma_x \otimes \sigma_z \otimes I$ \\
\hline
$|\eta_{12}\rangle$ 	 &$|+\rangle$	&$\alpha|111\rangle+\beta|100\rangle+\gamma|011\rangle+\delta|000\rangle$	&$\sigma_x \otimes \sigma_x \otimes \sigma_x $\\
\hline
$|\eta_{12}\rangle$ 	 &$|-\rangle$	&$\alpha|111\rangle+\beta|100\rangle-\gamma|011\rangle-\delta|000\rangle$	&$i\sigma_y \otimes \sigma_x \otimes \sigma_x $\\
\hline
$|\eta_{13}\rangle$ 	 &$|+\rangle$	&$\alpha|111\rangle-\beta|100\rangle+\gamma|011\rangle-\delta|000\rangle$	&$\sigma_x \otimes i\sigma_y \otimes \sigma_x$\\
\hline
$|\eta_{13}\rangle$ 	 &$|-\rangle$	&$\alpha|111\rangle-\beta|100\rangle-\gamma|011\rangle+\delta|000\rangle$	&$i\sigma_y \otimes i\sigma_y \otimes \sigma_x$\\
\hline
$|\eta_{14}\rangle$ 	 &$|+\rangle$	&$\alpha|111\rangle+\beta|100\rangle-\gamma|011\rangle-\delta|000\rangle$	&$i\sigma_y \otimes \sigma_x \otimes \sigma_x$\\
\hline
$|\eta_{14}\rangle$ 	 &$|-\rangle$	&$\alpha|111\rangle+\beta|100\rangle+\gamma|011\rangle+\delta|000\rangle$	&$\sigma_x \otimes \sigma_x \otimes \sigma_x$\\
\hline
$|\eta_{15}\rangle$ 	 &$|+\rangle$	&$\alpha|111\rangle-\beta|100\rangle-\gamma|011\rangle+\delta|000\rangle$	&$i\sigma_y \otimes i\sigma_y \otimes \sigma_x $\\
\hline
$|\eta_{15}\rangle$ 	 &$|-\rangle$	&$\alpha|111\rangle-\beta|100\rangle+\gamma|011\rangle-\delta|000\rangle$	&$\sigma_x \otimes i\sigma_y \otimes \sigma_x $\\
\hline
\end{tabular}}
\label{tab1}
\end{table}

There is another non-equivalent distribution of particles among the three parties. In this case, of the first $|\Psi_{\text{GHZ}}^0\rangle$ state, particle 1 belongs to Alice and rest belongs to Bob and of the the second $|\Psi_{\text{GHZ}}^0\rangle$ state, Alice has particle 1, Bob has particle 2 and Charlie has particle 3. The quantum channel in this case (in AABBBC format) is given by
\begin{equation*}
|\tilde{\Psi}_G\rangle =\frac{1}{2}[|000000\rangle+|010011\rangle+|101100\rangle+|111111\rangle].
\end{equation*}
The corresponding three qubit state (in Alice's possession) used for QIS is 
$|\tilde{\xi}\rangle=(\alpha|000\rangle+\beta|001\rangle+\gamma|110\rangle+\delta|111\rangle)$ with $|\alpha|^2+|\beta|^2+|\gamma|^2+|\delta|^2=1$. The combined state is given by
{\small
\begin{equation*}
|\tilde{\xi}\rangle\otimes|\tilde{\Psi}_G\rangle = \frac{1}{2}[\alpha|000\rangle+\beta|011\rangle+\gamma|100\rangle+\delta|111\rangle]\otimes 
[|000000\rangle+|010011\rangle+|101100\rangle+|111111\rangle].
\end{equation*}
}
In this case, Alice performs a 5-qubit measurement in the orthonormal basis
$\tilde{\eta}_i$, $i = 0, \ldots 15$ as given below.
\begin{equation*}
|\tilde{\eta}_{i}\rangle  =  \frac{1}{2}(\sigma_x^{4})^{i_3}(\sigma_x^{5})^{i_2}[|00000\rangle + (-1)^{i_1} |00101\rangle 
 +(-1)^{i_0} |11010\rangle + (-)^{i_1+i_0}|11111\rangle],
\end{equation*}
where $i_3i_2i_1i_0$ is the 4-bit binary representation of $i$ and 
$\sigma_x^b$ denotes $\sigma_x$ operated on the $b$-th bit ($b=1, \ldots, 5$)
from the left end corresponding to each term. Alternatively, starting with
\begin{equation*}
|\tilde{\eta}_{0}\rangle=\frac{1}{2}[|00000\rangle+|00101\rangle+|11010\rangle+|11111\rangle],
\end{equation*}
we can represent the other states in terms of $|\tilde{\eta}_{0}\rangle$ as
\begin{equation*}
|\tilde{\eta}_{i}\rangle = (\sigma_x^{4})^{i_3}(\sigma_x^{5})^{i_2} (\sigma_z^{4})^{i_1}(\sigma_z^{5})^{i_0}) |\tilde{\eta}_{0}\rangle.
\end{equation*}

Alice communicates her measurement results via four classical bits. Charlie communicates his measurement result in the Hadamard basis through one cbit. Alice and Charlie's measurement results, their communicated results to Bob and Bob's corresponding operations are listed in Table~\ref{tab2}. Similar to 
Table~\ref{tab1}, here also Alice's information corresponding
to the state $\tilde{\eta}_i$ is given by the 4-bit binary representation of $i$. Further, Charlie's information is 0 or 1 depending on whether his state is
$|+\rangle$ or $|-\rangle$.

\begin{table}[hb]
\tcaption{Strategy for recovering the three-qubit state}
\centerline{\footnotesize\smalllineskip
\begin{tabular}{c c c c}\\
\hline
Alice's    	&Charlie's  &Bob's	&Bob's\\
state 	 	&state 	    &State 	&Operation\\
\hline
$|\tilde{\eta}_0\rangle$ 	 &$|+\rangle$	&$\alpha|000\rangle+\beta|001\rangle+\gamma|110\rangle+\delta|111\rangle$ &$I \otimes I\otimes I$ \\
\hline
$|\tilde{\eta}_0\rangle$ 	 &$|-\rangle$	&$\alpha|000\rangle+\beta|001\rangle-\gamma|110\rangle-\delta|111\rangle$ &$I \otimes \sigma_z\otimes I$ \\
\hline
$|\tilde{\eta}_1\rangle$ 	 &$|+\rangle$	&$\alpha|000\rangle-\beta|001\rangle+\gamma|110\rangle-\delta|111\rangle$ &$I \otimes I \otimes \sigma_z $ \\
\hline
$|\tilde{\eta}_1\rangle$ 	 &$|-\rangle$	&$\alpha|000\rangle-\beta|001\rangle-\gamma|110\rangle+\delta|111\rangle$ &$I \otimes \sigma_z\otimes \sigma_z $ \\
\hline
$|\tilde{\eta}_2\rangle$ 	 &$|+\rangle$	&$\alpha|000\rangle+\beta|001\rangle-\gamma|110\rangle-\delta|111\rangle$ &$I \otimes \sigma_z \otimes I$ \\
\hline
$|\tilde{\eta}_2\rangle$ 	 &$|-\rangle$	&$\alpha|000\rangle+\beta|001\rangle+\gamma|110\rangle+\delta|111\rangle$ &$I \otimes I \otimes I$ \\
\hline
$|\tilde{\eta}_3\rangle$ 	 &$|+\rangle$	&$\alpha|000\rangle-\beta|001\rangle-\gamma|110\rangle+\delta|111\rangle$ &$I \otimes \sigma_z \otimes \sigma_z$ \\
\hline
$|\tilde{\eta}_3\rangle$ 	 &$|-\rangle$	&$\alpha|000\rangle-\beta|001\rangle+\gamma|110\rangle-\delta|111\rangle$ &$I \otimes I \otimes \sigma_z$ \\
\hline
$|\tilde{\eta}_4\rangle$ 	 &$|+\rangle$	&$\alpha|001\rangle+\beta|000\rangle+\gamma|111\rangle+\delta|110\rangle$ &$I \otimes I \otimes \sigma_x $ \\
\hline
$|\tilde{\eta}_4\rangle$ 	 &$|-\rangle$	&$\alpha|001\rangle+\beta|000\rangle-\gamma|111\rangle-\delta|110\rangle$ &$I \otimes \sigma_z \otimes \sigma_x $ \\
\hline
$|\tilde{\eta}_5\rangle$ 	 &$|+\rangle$	&$\alpha|001\rangle-\beta|000\rangle+\gamma|111\rangle-\delta|110\rangle$ &$I \otimes I \otimes i\sigma_y $ \\
\hline
$|\tilde{\eta}_5\rangle$ 	 &$|-\rangle$	&$\alpha|001\rangle-\beta|000\rangle-\gamma|111\rangle+\delta|110\rangle$ &$I \otimes \sigma_z \otimes i\sigma_y $ \\
\hline
$|\tilde{\eta}_6\rangle$ 	 &$|+\rangle$	&$\alpha|001\rangle+\beta|000\rangle-\gamma|111\rangle-\delta|110\rangle$ &$I \otimes \sigma_z \otimes \sigma_x $ \\
\hline
$|\tilde{\eta}_6\rangle$ 	 &$|-\rangle$	&$\alpha|001\rangle+\beta|000\rangle+\gamma|111\rangle+\delta|110\rangle$ &$I \otimes I \otimes \sigma_x $ \\
\hline
$|\tilde{\eta}_7\rangle$ 	 &$|+\rangle$	&$\alpha|001\rangle-\beta|000\rangle-\gamma|111\rangle+\delta|110\rangle$ &$I \otimes \sigma_z \otimes i\sigma_y $ \\
\hline
$|\tilde{\eta}_7\rangle$ 	 &$|-\rangle$	&$\alpha|001\rangle-\beta|000\rangle+\gamma|111\rangle-\delta|110\rangle$ &$I \otimes I \otimes i\sigma_y $ \\
\hline
$|\tilde{\eta}_8\rangle$ 	 &$|+\rangle$	&$\alpha|110\rangle+\beta|111\rangle+\gamma|000\rangle+\delta|001\rangle$ &$\sigma_x \otimes \sigma_x \otimes I$ \\
\hline
$|\tilde{\eta}_8\rangle$ 	 &$|-\rangle$	&$\alpha|110\rangle+\beta|111\rangle-\gamma|000\rangle-\delta|001\rangle$ &$\sigma_x \otimes i\sigma_y \otimes I$ \\
\hline
$|\tilde{\eta}_9\rangle$ 	 &$|+\rangle$	&$\alpha|110\rangle-\beta|111\rangle+\gamma|000\rangle-\delta|001\rangle$ &$\sigma_x \otimes \sigma_x \otimes \sigma_z$ \\
\hline
$|\tilde{\eta}_9\rangle$ 	 &$|-\rangle$	&$\alpha|110\rangle-\beta|111\rangle-\gamma|000\rangle+\delta|001\rangle$ &$\sigma_x \otimes i\sigma_y \otimes \sigma_z$ \\
\hline
$|\tilde{\eta}_{10}\rangle$ 	 &$|+\rangle$	&$\alpha|110\rangle+\beta|111\rangle-\gamma|000\rangle-\delta|001\rangle$ &$\sigma_x \otimes i\sigma_y \otimes I$ \\
\hline
$|\tilde{\eta}_{10}\rangle$ 	 &$|-\rangle$	&$\alpha|110\rangle+\beta|111\rangle+\gamma|000\rangle+\delta|001\rangle$ &$\sigma_x \otimes \sigma_x \otimes I$ \\
\hline
$|\tilde{\eta}_{11}\rangle$ 	 &$|+\rangle$	&$\alpha|110\rangle-\beta|111\rangle-\gamma|000\rangle+\delta|001\rangle$ &$\sigma_x \otimes i\sigma_y \otimes \sigma_z$ \\
\hline
$|\tilde{\eta}_{11}\rangle$ 	 &$|-\rangle$	&$\alpha|110\rangle-\beta|111\rangle+\gamma|000\rangle-\delta|001\rangle$ &$\sigma_x \otimes \sigma_x \otimes \sigma_z$ \\
\hline
$|\tilde{\eta}_{12}\rangle$ 	 &$|+\rangle$	&$\alpha|111\rangle+\beta|110\rangle+\gamma|001\rangle+\delta|000\rangle$	&$\sigma_x \otimes \sigma_x \otimes \sigma_x $\\
\hline
$|\tilde{\eta}_{12}\rangle$ 	 &$|-\rangle$	&$\alpha|111\rangle+\beta|110\rangle-\gamma|001\rangle-\delta|000\rangle$	&$\sigma_x \otimes i\sigma_y \otimes \sigma_x $\\
\hline
$|\tilde{\eta}_{13}\rangle$ 	 &$|+\rangle$	&$\alpha|111\rangle-\beta|110\rangle+\gamma|001\rangle-\delta|000\rangle$	&$\sigma_x \otimes \sigma_x \otimes i\sigma_y$\\
\hline
$|\tilde{\eta}_{13}\rangle$ 	 &$|-\rangle$	&$\alpha|111\rangle-\beta|110\rangle-\gamma|001\rangle+\delta|000\rangle$	&$\sigma_x \otimes i\sigma_y \otimes i\sigma_y$\\
\hline
$|\tilde{\eta}_{14}\rangle$ 	 &$|+\rangle$	&$\alpha|111\rangle+\beta|110\rangle-\gamma|001\rangle-\delta|000\rangle$	&$\sigma_x \otimes i\sigma_y \otimes \sigma_x$\\
\hline
$|\tilde{\eta}_{14}\rangle$ 	 &$|-\rangle$	&$\alpha|111\rangle+\beta|110\rangle+\gamma|001\rangle+\delta|000\rangle$	&$\sigma_x \otimes \sigma_x \otimes \sigma_x$\\
\hline
$|\tilde{\eta}_{15}\rangle$ 	 &$|+\rangle$	&$\alpha|111\rangle-\beta|110\rangle-\gamma|001\rangle+\delta|000\rangle$	&$\sigma_x \otimes i\sigma_y \otimes i\sigma_y $\\
\hline
$|\tilde{\eta}_{15}\rangle$ 	 &$|-\rangle$	&$\alpha|111\rangle-\beta|110\rangle+\gamma|001\rangle-\delta|000\rangle$	&$\sigma_x \otimes \sigma_x \otimes i\sigma_y $\\
\hline
\end{tabular}}
\label{tab2}
\end{table}

\section{\label{sec:4QIS}Quantum Information Splitting of a Four-qubit State}
In this case, the quantum channel connecting Alice, Bob and Charlie 
(in ABBBBC format) is given by $\frac{1}{2}[|000000\rangle+|000111\rangle+|111000\rangle+|111111\rangle]$. Here, Alice has particle 1, Bob has particles 2,3,4,5 and Charlie has particle 6. The four-qubit state (in Alice's possession) used for QIS is $|\zeta\rangle=\alpha(|0000\rangle +|0011\rangle)+\beta(|1100\rangle+|1111\rangle)$ with $|\alpha|^2+|\beta|^2 =\frac{1}{2}$. Alice combines this state with the qubit of the shared channel and performs
a 5-qubit measurement on the following joint state of the system:
\begin{eqnarray*}
|\zeta\rangle\otimes|\Psi_G\rangle
& = & [\alpha(|0000\rangle+|0011\rangle)+\beta(|1100\rangle+|1111\rangle)]\\
& & \otimes
\frac{1}{2}[|000000\rangle+|000111\rangle +|111000\rangle+|111111\rangle]\\
& = & \frac{\alpha}{2}[(|\nu_0\rangle+|\nu_1\rangle)(|00000\rangle+|00111\rangle)] +(|\nu_2\rangle+|\nu_3\rangle)(|11000\rangle+|11111\rangle)\\
& & +\frac{\beta}{2}[(|\nu_2\rangle-|\nu_3\rangle)(|00000\rangle+|00111\rangle)]
 +(|\nu_0\rangle-|\nu_1\rangle)(|11000\rangle+|11111\rangle),
\end{eqnarray*}

where $\nu_i$'s are Alice's five-qubit measurement basis given by
\begin{equation*}
\begin{split}
|\nu_{0}\rangle=\frac{1}{2}[|00000\rangle+|00110\rangle+|11001\rangle+|11111\rangle],\\
|\nu_{1}\rangle=\frac{1}{2}[|00000\rangle+|00110\rangle-|11001\rangle-|11111\rangle],\\
|\nu_{2}\rangle=\frac{1}{2}[|00001\rangle+|00111\rangle+|11000\rangle+|11110\rangle],\\
|\nu_{3}\rangle=\frac{1}{2}[|00001\rangle+|00111\rangle+|11000\rangle+|11110\rangle].\\
\end{split}
\end{equation*}
Alice and Charlie's measurement results, corresponding classical communication and Bob's unitary operations are listed in Table~\ref{tab3}.
In our notation, Alice's information corresponding
to the state $\nu_i$ is given by the 4-bit binary representation of $i$
and Charlie's information is 0 or 1 depending on whether his state is
$|+\rangle$ or $|-\rangle$.

\begin{table*}[hb]
\tcaption{Strategy for recovering the four-qubit state}
\centerline{\footnotesize\smalllineskip
\begin{tabular}{c c c c}\\
\hline
Alice's 	&Charlie's      &Bob's	&Bob's\\
state 	  	&state 		&State 	&Operation\\
\hline
$|\nu_{0}\rangle$ 	 &$|+\rangle$ 	&$\alpha(|0000\rangle+|0011\rangle)+\beta(|1100\rangle+|1111\rangle)$		& $I \otimes I\otimes I\otimes I$ \\
\hline
$|\nu_{0}\rangle$ 	 &$|-\rangle$ 	&$\alpha(|0000\rangle-|0011\rangle)+\beta(|1100\rangle-|1111\rangle)$		& $I \otimes I \otimes \sigma_z\otimes I$ \\
\hline
$|\nu_{1}\rangle$ 	 &$|+\rangle$ 	&$\alpha(|0000\rangle+|0011\rangle)-\beta(|1100\rangle+|1111\rangle)$		& $I \otimes \sigma_z\otimes I\otimes I$ \\
\hline
$|\nu_{1}\rangle$ 	 &$|-\rangle$ 	&$\alpha(|0000\rangle-|0011\rangle)-\beta(|1100\rangle-|1111\rangle)$		& $I \otimes \sigma_z\otimes \sigma_z\otimes I$ \\
\hline
$|\nu_{2}\rangle$ 	 &$|+\rangle$ 	&$\alpha(|1100\rangle+|1111\rangle)+\beta(|0000\rangle+|0011\rangle)$		& $\sigma_x\otimes \sigma_x\otimes I\otimes I$ \\
\hline
$|\nu_{2}\rangle$ 	 &$|-\rangle$ 	&$\alpha(|1100\rangle-|1111\rangle)+\beta(|0000\rangle-|0011\rangle)$		& $\sigma_x\otimes \sigma_x\otimes \sigma_z\otimes I$ \\
\hline
$|\nu_{3}\rangle$ 	 &$|+\rangle$ 	&$\alpha(|1100\rangle+|1111\rangle)-\beta(|0000\rangle+|0011\rangle)$		& $\sigma_x\otimes i\sigma_y\otimes I\otimes I$ \\
\hline
$|\nu_{3}\rangle$ 	 &$|-\rangle$ 	&$\alpha(|1100\rangle-|1111\rangle)-\beta(|0000\rangle-|0011\rangle)$		& $\sigma_x\otimes i\sigma_y\otimes \sigma_z\otimes I$ \\
\hline
\end{tabular}}
\label{tab3}
\end{table*}

\section{\label{sec:conc}Conclusion}
Since a single arbitrary qubit can be split successfully among three parties using a GHZ state, it is quite obvious that one can split an arbitrary two-qubit states among three parties using two GHZ states. But splitting a three qubit 
state, or even a special form of it, is not very obvious. In this article, we have shown that the restricted class of two non-equivalent types of three-qubit states can be used for QIS among three parties using a pair of GHZ states as the quantum channel. We also described a protocol for QIS of a restricted class of four-qubit states among the parties using the same channel. Though the realizations of various quantum processing tasks involving single and two-qubit states using various multipartite systems as resource, both theoretically and experimentally, are well understood, there is still some lack of knowledge in our understanding in multi-qubit states. We hope this article will enhance our knowledge in that direction.

\section*{Acknowledgments}
KN would like to thank the University Grants Commission of India, New Delhi for partially supporting this research under Faculty Development Programme [Grant No. F. TF.WC2-005-04/08-09 (ERO)].

\end{document}